\documentclass[manuscript]{aastex}
\usepackage{longtable}

\shorttitle{Automatic Classification of SDSS stellar spectra}

\begin{document}

%% LaTeX will automatically break titles if they run longer than
%% one line. However, you may use \\ to force a line break if
%% you desire.

\title{Automated Classification of Sloan Digital Sky Survey (SDSS)
Stellar Spectra using Artificial Neural Networks}

%% Use \author, \affil, and the \and command to format
%% author and affiliation information.
%% Note that \email has replaced the old \authoremail command
%% from AASTeX v4.0. You can use \email to mark an email address
%% anywhere in the paper, not just in the front matter.
%% As in the title, use \\ to force line breaks.

\author{Mahdi Bazarghan\altaffilmark{1} and Ranjan Gupta\altaffilmark{1}}
\affil{Inter University Center for Astronomy and Astrophysics}
%% Notice that each of these authors has alternate affiliations, which
%% are identified by the \altaffilmark after each name.  Specify alternate
%% affiliation information with \altaffiltext, with one command per each
%% affiliation.
\altaffiltext{1}{Inter University Center for Astronomy and
Astrophysics, Post Bag 4, Ganeshkhind, Pune-411007, India}
\email{mahdi@iucaa.ernet.in}
%% Mark off your abstract in the ``abstract'' environment. In the manuscript
%% style, abstract will output a Received/Accepted line after the
%% title and affiliation information. No date will appear since the author
%% does not have this information. The dates will be filled in by the
%% editorial office after submission.

\begin{abstract}
Automated techniques have been developed to automate the process of
classification of objects or their analysis. The large datasets
provided by upcoming spectroscopic surveys with dedicated telescopes
urges scientists to use these automated techniques for analysis of
such large datasets which are now available to the community. Sloan
Digital Sky Survey (SDSS) is one of such surveys releasing massive
datasets. We use Probabilistic Neural Network (PNN) for automatic
classification of about 5000 SDSS spectra into 158 spectral type of
a reference library ranging from O type to M type stars.
\end{abstract}

%% Keywords should appear after the \end{abstract} command. The uncommented
%% example has been keyed in ApJ style. See the instructions to authors
%% for the journal to which you are submitting your paper to determine
%% what keyword punctuation is appropriate.

\keywords{Probabilistic Neural Network, Spectra, Sloan Digital Sky
Survey, Spectral classification}

%% From the front matter, we move on to the body of the paper.
%% In the first two sections, notice the use of the natbib \citep
%% and \citet commands to identify citations.  The citations are
%% tied to the reference list via symbolic KEYs. The KEY corresponds
%% to the KEY in the \bibitem in the reference list below. We have
%% chosen the first three characters of the first author's name plus
%% the last two numeral of the year of publication as our KEY for
%% each reference.

%% Authors who wish to have the most important objects in their paper
%% linked in the electronic edition to a data center may do so by tagging
%% their objects with \objectname{} or \object{}.  Each macro takes the
%% object name as its required argument. The optional, square-bracket
%% argument should be used in cases where the data center identification
%% differs from what is to be printed in the paper.  The text appearing
%% in curly braces is what will appear in print in the published paper.
%% If the object name is recognized by the data centers, it will be linked
%% in the electronic edition to the object data available at the data centers
%%
%% Note that for sources with brackets in their names, e.g. [WEG2004] 14h-090,
%% the brackets must be escaped with backslashes when used in the first
%% square-bracket argument, for instance, \object[\[WEG2004\] 14h-090]{90}).
%%  Otherwise, LaTeX will issue an error.

\section{Introduction}

Intelligent systems based on pattern recognition tools like
Artificial Neural Networks (ANNs) are now been used in various
fields in applications like time series based event predictions,
object classifications, data compression etc. In the past decade
ANNs have found applications in classification of spectra,
star-galaxy etc which are typical Astronomical requirements where
large data bases are now coming up. The Sloan Digital Sky Survey
\citep{b20} is one of such publicly available source of data where
the need of automatic classification of a large data set provides an
ideal ground for ANNs. Previous studies involving spectral
classification using ANNs include: \citet{b7,b9}, \citet{b18},
\citet{b19}, \citet{b14,b15}, \citet{b10} and also automated
analysis of stellar spectra by \citet{Allende}. These pioneering
efforts resulted in setting up the scene where ANNs have proven to
be an established tool for classification of large set of stellar
spectra.

We have used Probabilistic Neural Network \citep{b4} in MATLAB
programming to classify 4999 test spectra of the SDSS into a set of
158 training spectra from a reference spectral library \citep{b11}.
%%%%%%%%%%%%%%%%%%%%%%Inja ezafe shode %%%%%%%%%%%%%%%%%%%
Probabilistic neural network is intrinsically a classifier with four
layer network. Input layer is fully connected to the next layer that
is the pattern layer. There is one pattern node for each training
example in this layer. The next layer is a summation layer which sums
the inputs from pattern units. The output layer with one neuron
represents the maximum value in the summation layer. PNN uses a
supervised training set to develop probability density functions
within a pattern layer. As new pattern vectors are presented to the
PNN for classification, they are serially propagated through the
hidden layer by computing the dot product between the new pattern
and each pattern stored in the hidden layer.
%%%%%%%%%%%%%%%%%%%%%%%%%

The result of classification of about 5000 SDSS stellar spectra is
possibly a first attempt on this data set, though ANNs are already
used on SDSS data for finding Galaxy types \citep{b12}, measuring
photometric redshifts \citep{b17} and separation of stars and
galaxies \citep{b13}.

The section 2 describes the SDSS spectral data set; section 3 and 4
describe the PNN tool which has been applied to the data set and the
classification process details for this work and finally the
sections 5 and 6 give the results and discussion.

\section{SDSS spectral data set and reference set of spectra}
%% In a manner similar to \objectname authors can provide links to dataset
%% hosted at participating data centers via the \dataset{} command.  The
%% second curly bracket argument is printed in the text while the first
%% parentheses argument serves as the valid data set identifier.  Large
%% lists of data set are best provided in a table (see Table 3 for an example).
%% Valid data set identifiers should be obtained from the data center that
%% is currently hosting the data.
%%
%% Note that AASTeX interprets everything between the curly braces in the
%% macro as regular text, so any special characters, e.g. "#" or "_," must be
%% preceded by a backslash. Otherwise, you will get a LaTeX error when you
%% compile your manuscript.  Special characters do not
%% need to be escaped in the optional, square-bracket argument.

Sloan Digital Sky Survey is the most ambitious astronomical survey
project ever undertaken. The survey will map in detail one-quarter
of the entire sky with five broad band filters, determining the
positions and absolute brightness of more than 100 million celestial
objects. A technical summary of the survey is given in \citet{b20}.
The data release to the community so far consists of the June 2001
Early Data Release \citep{b16}, the April 2003 Data Release One
\citep{b1}, the March 2004 Data Release Two (DR2) \citep{b2}, the
September 2004 Data Release Three \citep{b3}, the June 2005 Data
Release Four \citep{b21}, the June 2006 Data Release Five
\citep{b22} and the June 2007 Data Release six \citep{b23}.

In order to have successful application of the ANN techniques and
achieving effective results, one has to prepare the dataset well
before applying to the neural network. They must be all uniform,
having same wavelength scale, the starting and end wavelengths must
be same for all the spectra and they must also have closely match
spectral resolution. These must be valid to both the training and
test dataset.

The training set for the PNN is a set of 158 spectra taken from
\citet{b11} (hereafter JHC which contains a total 161 spectra),
which covers the wavelength range of 3510-7427 \AA~ for various O to
M type stars. The set of 4999 spectra from SDSS with wavelength
range of 3800-9200 \AA~ forms the test set which gets classified
into 158 spectro-luminosity classes of the reference JHC library.
The SDSS test data were in fits format with six binary tables
attached to the images. %Using the IRAF task {\it wspectext}, we
%converted the spectra to a text format (two columns of wavelength
%and fluxes) which is acceptable to the artificial neural networks.
We converted the fits images into ASCII files (i.e. two columns of
wavelength and fluxes) which are acceptable to the artificial neural
networks, using the IRAF task {\it wspectext}. The spectral
resolution of JHC is FWHM=4.5 \AA~ with one flux value per 1.4 \AA~
and the SDSS has a resolution of about FWHM=3.25 \AA~ with each flux
value at 0.5 \AA~. The spectral resolution of testing data set is
brought to 4.5 \AA~ with sampling rate of 5 \AA~, thus the final
test and train sets are re-binned at 5 \AA~ steps resulting in 561
data points for each spectra in training and test datasets. This is
achieved by using appropriate convolution and spline fitting
routines. Also both the libraries are normalized into unity.
%%%%%%%%%%%%%%%%%%%%%%%%%%%%%%%%%%%%%%%%%%%%%%%%%%%%%%%

%% In this section, we use  the \subsection command to set off
%% a subsection.  \footnote is used to insert a footnote to the text.

%% Observe the use of the LaTeX \label
%% command after the \subsection to give a symbolic KEY to the
%% subsection for cross-referencing in a \ref command.
%% You can use LaTeX's \ref and \label commands to keep track of
%% cross-references to sections, equations, tables, and figures.
%% That way, if you change the order of any elements, LaTeX will
%% automatically renumber them.

%% This section also includes several of the displayed math environments
%% mentioned in the Author Guide.
%########################################

%\footnote{Footnotes can be inserted like this.}

%########################################
\section{Probabilistic Neural Networks}

The Probabilistic Neural Network (PNN) was developed by \citet{b4}.
This network provides a general solution to pattern classification
problems by following an approach developed in statistics, called
Bayesian classifiers. This technique is constructed using ideas from
classical probability theory, such as Bayesian classification and
classical estimators for probability density functions (pdf)
\citep{b6}, to form a neural network for pattern classification and
estimation of class membership \citep{b5}. The Figure 1 shows the
architecture of PNN. It is a four layer feedforward network
consisting of input layer, pattern layer, summation layer and the
output layer. The input layer contains n nodes to accept an
n-dimensional feature vector (n=561) and is fully connected to the
pattern layer which passes the input into pattern layer. The pattern
layer consists of K groups of pattern nodes. The $k^{th}$ group in
the pattern layer contains $N_k$ number of pattern nodes, where, K
is the number of training patterns or classes (i.e. 158 JHC
library). The summation layer is having K nodes, one node for each
class in the pattern layer. Pattern nodes of each $k^{th}$ group in
the pattern layer are connected to the corresponding $k^{th}$
summation node in the summation layer.
%If we consider the
%input vector $X = (x_1, x_2, ..., x_n ) \in \Re^n$, applied to n
%input neurons, the neurons of the pattern layer are divided into K
%groups and each group presents one training pattern here is 158
%training patterns of JHC library.
%%%%%%%%%%%%%%%%% az inja ezafe shode %%%%%%%%%%%%%%
The probabilistic neural network uses the following estimator for
the probability density function of $k^{th}$ group:

\begin{equation}
S_k (X) = \frac{1}{(2 \pi
\sigma^2)^{n/2}}\frac{1}{N_k}\sum_{i=1}^{N_k}\exp \left( -
\frac{\parallel X - X_{k,i} \parallel^2}{2 \sigma^2} \right).
\end{equation}
%%%%%%%%%%%%%%%%%%% ta inja %%%%%%%%%%%%%%%%%%%%%%%%
where $X_{k,i} \in \Re^n$ is the center of the kernel, and $\sigma$
also known as the spread or smoothing parameter which is the
deviation of the Gaussian function. Finally at the output layer the
pattern vector X will be classified as a class which corresponds to
the summation unit with maximum value,
\begin{equation}
C(X) = arg\max_{1 \le k \le K}( S_k ).
%C(X)=\rm arg \ max\it(_{\substack{\\ \\ \hskip -0.8 cm 1 \le k \le
%K}} S_k).
\end{equation}

%########################################
\section{Classification}

The application of neural networks to classification problems is
conceptually the most consistent with their structure and function.
Considering a finite set of states or classes, the objective in
classification applications are the assignment of a random samples
to one of those states with minimum probability of errors. Each
sample is described by a set of parameters which form a vector,
usually refereed to as the feature vector. The development of such a
classification system can be achieved by appropriately training a
neural network in such a way that it provides an output
corresponding to one of the classes, provided that, the training
sample used in forming its inputs belong to this class. The ability
of the neural network to correctly classify a test sample that is
close in some sense to one of the training samples, related directly
to its generalization ability.

Stellar spectra classification is one of the application areas of
the artificial neural networks. Here we use PNN for the
classification, an excellent classifier with very fast training
process and no local minima issues which outperforms other
classifiers including back-propagation. After preprocessing both the
training and test datasets, the training set with 158 spectra and
561 flux bins each, are given to the network for the training. So in
the input layer of the network there will be 561 number of nodes
corresponding to dimension of the spectra. Then the input will be
passed to the next layer i.e. pattern layer. All the 158 training
samples will be stored in the pattern layer and this layer organizes
as groups and each group will be dedicated to one class of spectra.
Hence there will be 158 groups in the pattern layer, each containing
as many neurons as the number of flux bins in the spectra i.e. 561.
In the testing stage, 4999 test spectra from SDSS will be given to
the input layer and the distance between them and the training input
vectors will be evaluated and then a vector will be produced by
which closeness and similarity of the test data and training data
can be judged. Finally these vectors will be summed up in the
summation layer to produce an output as vector of probabilities and
the maximum probability will be selected at the output node. The
neuron corresponding to the highest probability value will be the
closest class to the test spectra, for example if neuron $S_{k}$ in
Figure 1 is having the highest output, it indicates that the given
input belongs to the $K^{th}$ group or class.
%########################################
%% The \notetoeditor{TEXT} command allows the author to communicate
%% information to the copy editor.  This information will appear as a
%% footnote on the printed copy for the manuscript style file.  Nothing will
%% appear on the printed copy if the preprint or
%% preprint2 style files are used.

%% The eqnarray environment produces multi-line display math. The end of
%% each line is marked with a \\. Lines will be numbered unless the \\
%% is preceded by a \nonumber command.
%% Alignment points are marked by ampersands (&). There should be two
%% ampersands (&) per line.

%################\notetoeditor{Figures 1 and 2 should appear side-by-side in print}
\section{Result}

Automated classification of the SDSS-DR2 set of 4999 spectra were
carried out using the PNN algorithm. Since we believe that this is a
first attempt of classifying the SDSS spectra of originally unknown
spectro-luminosity classes into a known set of classes of a JHC
reference library; we have provided the complete classification
result in Table 4.\footnote{The table is provided in separate file}
This table gives the name of star, Spectro-Luminosity class given by
ANN and Chi-square error of SDSS and its corresponding JHC reference
spectra.

To illustrate the quality of fits of the SDSS spectra with respect
to the JHC spectra, we have plotted some typical spectro-luminosity
classes from each main spectral types in Figures 2-5 and 6-9. These
plots consist of 16 panels per page covering most of the O to M main
spectral types. The JHC reference spectrum are shown in bold black
and the SDSS spectra in thin blue lines. Tables 1 and 2 lists the
SDSS spectra name; corresponding JHC class obtained by the PNN and
the corresponding $\chi^2$ value of these plots.

\begin{table}
\begin{center}
\renewcommand{\arraystretch}{0.75}
%\begin{center}
\caption{List of spectra appearing in figures 2-5}
%\begin{center}
\begin{tabular}{c  c  c}

\hline
SDSS    & Spectro-luminosity  & $\chi^2$ \\
Spectra & class & value\\  \hline

spSpec-52367-0332-184   & O9V   & 0.00142028\\
spSpec-51984-0279-228   & O7.5V & 0.00042054\\
spSpec-51986-0294-423   & O6.5V & 0.00701563\\
spSpec-51942-0301-626   & O5.5V & 0.00029600\\
spSpec-51673-0313-519   & O5V   & 0.00029920\\
spSpec-52000-0335-478   & O6.5III & 0.00048330\\
spSpec-51633-0268-008   & O6.5III & 0.00042562\\
spSpec-51941-0272-510   & B1.5V & 0.00052162\\
spSpec-51909-0276-370   & B3V   & 0.00062574\\
spSpec-51908-0277-521   & B4V   & 0.00202127\\
spSpec-51689-0312-031   & B6V   & 0.00089164\\
spSpec-51900-0278-242   & B8V   & 0.00078749\\
spSpec-51615-0303-060   & B1III & 0.00133665\\
spSpec-51928-0291-105   & B2III & 0.00027330\\
spSpec-51665-0311-575   & B2.5III & 0.00082641\\
spSpec-51990-0340-352   & B3III & 0.00076806\\
spSpec-51658-0282-110   & B4III & 0.00022405\\
spSpec-51691-0350-366   & B5III & 0.00030136\\
spSpec-51658-0282-067   & B7III & 0.00111983\\
spSpec-51792-0354-196   & B8III & 0.00076665\\
spSpec-51691-0342-435   & B9III & 0.00045657\\
spSpec-51663-0315-624   & B2II  & 0.00497184\\
spSpec-51818-0358-126   & B0.5I & 0.00156664\\
spSpec-52056-0329-201   & B1.5I & 0.00386845\\
spSpec-51688-0302-490   & B2I   & 0.00440111\\
spSpec-51908-0277-055   & B3I   & 0.00077680\\
spSpec-52370-0330-020   & B5I   & 0.00050732\\
spSpec-51990-0310-393   & B7I   & 0.00254126\\
spSpec-51671-0299-592   & B8I   & 0.00088153\\
spSpec-51994-0293-108   & B9I   & 0.00082505\\
spSpec-51789-0352-587   & B1I   & 0.00405424\\
spSpec-51792-0354-451   & A1V   & 0.00054832\\
spSpec-51994-0309-313   & A2V   & 0.00049425\\
spSpec-51789-0352-177   & A3V   & 0.00062889\\
spSpec-51990-0310-307   & A5V   & 0.00041014\\
\hline

\end{tabular}
\end{center}
\end{table}
%%%%%%%%%%%%%%%%%%%%%%%
\begin{table}
\renewcommand{\arraystretch}{0.75}
\caption{{\bf{Table 1}}(Contd.)}
\begin{center}
\begin{tabular}{c  c  c}

\hline

SDSS    & Spectro-luminosity  & $\chi^2$ \\
Spectra & class & value \\  \hline

spSpec-51658-0282-539   & A6V   & 0.00040037\\
spSpec-51630-0266-233   & A7V   & 0.00034089\\
spSpec-51994-0309-194   & A8V   & 0.00058785\\
spSpec-51691-0350-044   & A9V   & 0.00043968\\
spSpec-51994-0293-470   & A3III & 0.00072669\\
spSpec-51821-0359-263   & A6III & 0.00107148\\
spSpec-51821-0359-142   & A8III & 0.00043082\\
spSpec-51691-0342-518   & A0I   & 0.00054997\\
spSpec-51662-0308-083   & A1I   & 0.00183806\\
spSpec-51910-0269-055   & A2I   & 0.00128973\\
spSpec-51883-0271-121   & A3I   & 0.00061977\\
spSpec-51688-0302-167   & A4I   & 0.00156048\\
spSpec-51662-0308-503   & A7I   & 0.00258414\\
spSpec-51818-0358-078   & A9I   & 0.00064258\\
spSpec-51959-0283-021   & F0V   & 0.00046135\\
spSpec-51909-0276-426   & F3V   & 0.00048153\\
spSpec-51662-0308-343   & F4V   & 0.00047088\\
spSpec-51691-0342-357   & F5V   & 0.00066779\\
spSpec-51818-0358-060   & F6V   & 0.00057737\\
spSpec-52282-0328-287   & F7V   & 0.00129746\\
spSpec-52375-0326-035   & F8V   & 0.00240744\\
spSpec-52023-0287-188   & F9V   & 0.00105487\\
spSpec-51703-0353-570   & F0IV  & 0.00045788\\
spSpec-51699-0349-477   & F3IV  & 0.00054928\\
spSpec-51692-0339-051   & F0III & 0.00034837\\
spSpec-51910-0275-458   & F4III & 0.00054288\\
spSpec-51908-0277-315   & F5III & 0.00099659\\
spSpec-51692-0339-214   & F6III & 0.00098363\\
spSpec-51612-0280-593   & F7III & 0.00145028\\
\hline

\end{tabular}
\end{center}
\end{table}
%%%%%%%%%%%%%%%%%

\begin{table}
\renewcommand{\arraystretch}{0.75}
\caption{List of spectra appearing in the figures 6-9}
\begin{center}
\begin{tabular}{c  c  c  c }

\hline

SDSS    & Spectro-luminosity  & $\chi^2$\\
Spectra & class & value \\  \hline

spSpec-52294-0327-328   & F8III & 0.00202269\\
spSpec-51789-0352-129   & F5II  & 0.00080027\\
spSpec-51792-0354-184   & F0I   & 0.00086225\\
spSpec-51943-0300-544   & F2I   & 0.00091199\\
spSpec-51780-0351-300   & F3I   & 0.00235900\\
spSpec-51957-0304-551   & F4I   & 0.00157235\\
spSpec-52017-0366-297   & F5I   & 0.00569906\\
spSpec-51699-0349-570   & F8I   & 0.00085424\\
spSpec-51792-0354-388   & G0V   & 0.00106892\\
spSpec-51957-0304-589   & G1V   & 0.00119590\\
spSpec-51816-0360-178   & G2V   & 0.00070487\\
spSpec-51694-0338-614   & G3V   & 0.00074789\\
spSpec-51913-0274-130   & G4V   & 0.00097247\\
spSpec-51699-0349-117   & G6V   & 0.00077054\\
spSpec-51699-0349-252   & G7V   & 0.00065785\\
spSpec-51699-0349-139   & G9V   & 0.00247349\\
spSpec-51816-0360-287   & G2IV  & 0.00258696\\
spSpec-51816-0360-351   & G5IV  & 0.00152954\\
spSpec-51816-0360-568   & G0III & 0.00128494\\
spSpec-51780-0351-608   & G2III & 0.00176844\\
spSpec-51694-0338-196   & G4III & 0.00282095\\
spSpec-51994-0309-522   & G5III & 0.00172599\\
spSpec-51821-0359-556   & G6III & 0.00141486\\
spSpec-51699-0349-585   & G7III & 0.00202173\\
spSpec-51699-0349-441   & G8III & 0.00222946\\
spSpec-51689-0312-285   & G9III & 0.00181964\\
spSpec-51788-0355-169   & G8II  & 0.00260567\\
spSpec-51816-0360-100   & G9II  & 0.00265346\\
spSpec-51816-0360-497   & G0I   & 0.00359867\\
spSpec-51673-0313-125   & G1I   & 0.00501194\\
spSpec-51910-0269-313   & G3I   & 0.00302347\\
spSpec-51699-0349-340   & G5I   & 0.00416412\\
spSpec-51699-0349-328   & K0V   & 0.00130288\\
spSpec-51699-0349-522   & K4V   & 0.00200172\\
spSpec-51691-0342-037   & K5V   & 0.00259285\\
\hline

\end{tabular}
\end{center}
\end{table}
%%%%%%%%%%%%%%%
\begin{table}
\renewcommand{\arraystretch}{0.75}
\caption{{\bf{Table 2}}(Contd.)}
\begin{center}
\begin{tabular}{c  c  c  c }

\hline

SDSS    & Spectro-luminosity  & $\chi^2$\\
Spectra & class & value \\  \hline

spSpec-51780-0351-109   & K0III & 0.00156690\\
spSpec-51658-0282-282   & K2III & 0.00281755\\
spSpec-51780-0351-309   & K3III & 0.00319767\\
spSpec-51955-0298-293   & K4III & 0.00154753\\
spSpec-52313-0333-323   & K7III & 0.00637097\\
spSpec-51663-0315-540   & K6II  & 0.01685844\\
spSpec-51821-0359-591   & K0I   & 0.00183392\\
spSpec-51997-0337-129   & K5I   & 0.00271351\\
spSpec-51959-0283-455   & M0V   & 0.00648934\\
spSpec-51989-0363-264   & M1V   & 0.00333934\\
spSpec-51662-0308-320   & M5V   & 0.01469228\\
spSpec-52000-0288-064   & M3III & 0.01214894\\
spSpec-51658-0282-370   & M4III & 0.01844951\\
spSpec-51692-0339-064   & M3II  & 0.01805566\\
spSpec-52375-0326-135   & M1I   & 0.00607657\\
\hline

\end{tabular}
\end{center}
\end{table}

Figures 10 and 11 show a set of SDSS spectra which are having higher
$\chi^2$ values as compared to those of Figures 2-9; most of which
do not fit well with the corresponding JHC spectra obtained by the
PNN. Table 3 lists those set of spectra with high $\chi^2$ values.

\begin{table}
\renewcommand{\arraystretch}{0.75}
\caption{List of spectra with high $\chi^2$ values}
\begin{center}
\begin{tabular}{c  c  c  c }
\hline

SDSS    & Spectro-luminosity  & Chisq.\\
Spectra & class & value \\  \hline

spSpec-52017-0366-029   & O9V   & 0.00392584\\
spSpec-52375-0326-557   & O6.5III & 0.01137241\\
spSpec-51957-0304-067   & B4V   & 0.02226783\\
spSpec-52375-0326-565   & B7III & 0.01909759\\
spSpec-52000-0288-102   & B2III & 0.02392821\\
spSpec-52368-0331-206   & A7V   & 0.00735897\\
spSpec-51957-0273-141   & A8III & 0.01864975\\
spSpec-51789-0352-044   & A7I   & 0.01932068\\
spSpec-51662-0308-479   & F3V   & 0.00255453\\
spSpec-52056-0329-587   & F6III & 0.00313202\\
spSpec-51613-0305-192   & F4I   & 0.01137626\\
spSpec-51994-0309-410   & G4V   & 0.00792763\\
spSpec-52017-0366-310   & G2IV  & 0.01030766\\
spSpec-52017-0366-293   & G5I   & 0.01894512\\
spSpec-51818-0358-528   & K0V   & 0.00434149\\
spSpec-51673-0313-021   & K4III & 0.01364253\\
spSpec-51959-0283-365   & K0I   & 0.00883682\\
spSpec-51928-0291-599   & M1V   & 0.01796854\\
spSpec-51633-0268-613   & M3III & 0.01919349\\
spSpec-51942-0301-308   & M1I   & 0.02038125\\
spSpec-51999-0336-405   & O5V   & 0.83020251\\
spSpec-51816-0360-166   & O5V   & 0.02503208\\
\hline

\end{tabular}
\end{center}
\end{table}
%######################################
\section{Discussion}

As seen from classification result the high value of $\chi^2$
associated with the O5V spectro-luminosity class, two samples of
which are shown in Figure 11. Though there is one spectra in test
dataset which is correctly classified into this spectral type and
that is spSpec-51673-0313-519 with $\chi^2$ value of 0.00029920 and
is shown in Figure 2 which is well matching with its corresponding
JHC spectra; in Figures 10 and 11 we show the worst case of each
class with highest $\chi^2$ values.

A look at the panels in Figures 2-5 and 6-9 indicates that, except
for some spectra, most of them show excellent matches. If one
considers a $\chi^2$ value of 0.02 as a kind of limit, then, there
are about 600 spectra which has $\chi^2$ value worse than this limit
which corresponds to a success rate of about 88\%.

We could classify the stars based on their temperature and
luminosity classes with PNN technique, and only a few seconds were required to
classify all 4999 spectra (in the test session).
Considering this high speed of classification,
it would enable us to apply this technique to classify larger datasets of
the latest SDSS data releases. We are hoping to achieve better
results in the future classification of SDSS datasets by using
training samples from SDSS itself (i.e. the sample can be taken
from our classification
results by selecting the best match spectra of each spectral and
luminosity class having smallest $\chi^2$ value in this catalog).

This work also encourages to use ANN based techniques to classify
the complete SDSS set of spectra in near future.
%######################################
\section{Acknowledgments}

The first author thanks IUCAA and IASBS for providing the
computational facilities for this work.
%######################################

%######## Kheili nokteye jalebiye ############

%% The displaymath environment will produce the same sort of equation as
%% the equation environment, except that the equation will not be numbered
%% by LaTeX.
%#############################################

%################## FIGURES ###########################

\clearpage
\begin{figure}[h!]
\vspace{11pc}
\hskip -2 cm
\includegraphics[scale=0.75,angle=-90]{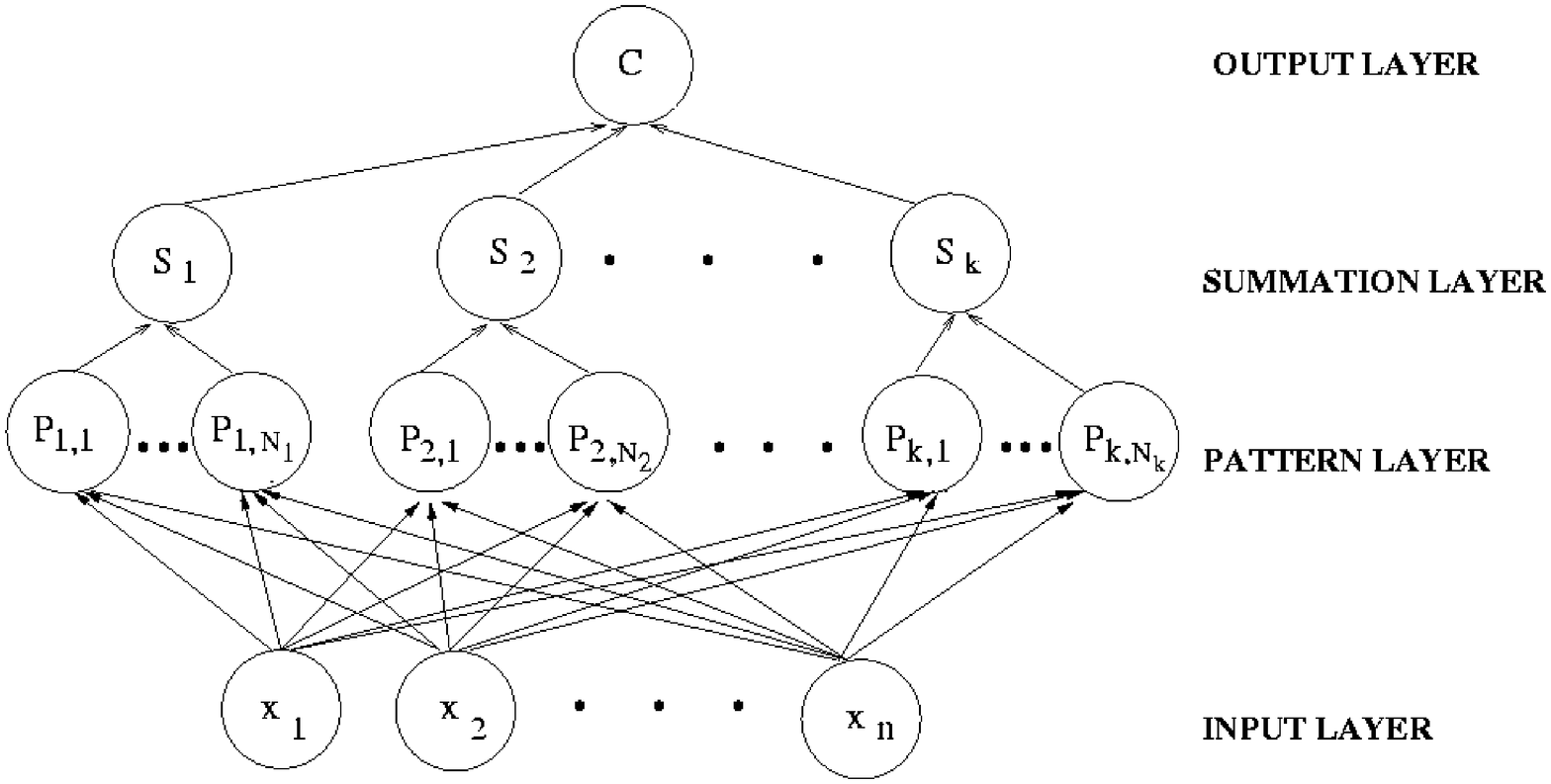}
\vspace{5pc}
\caption{Schematic of a typical Probabilistic Neural
Network.}
\end{figure}
%%%
\clearpage
\begin{figure}
\hskip -1 cm
\includegraphics{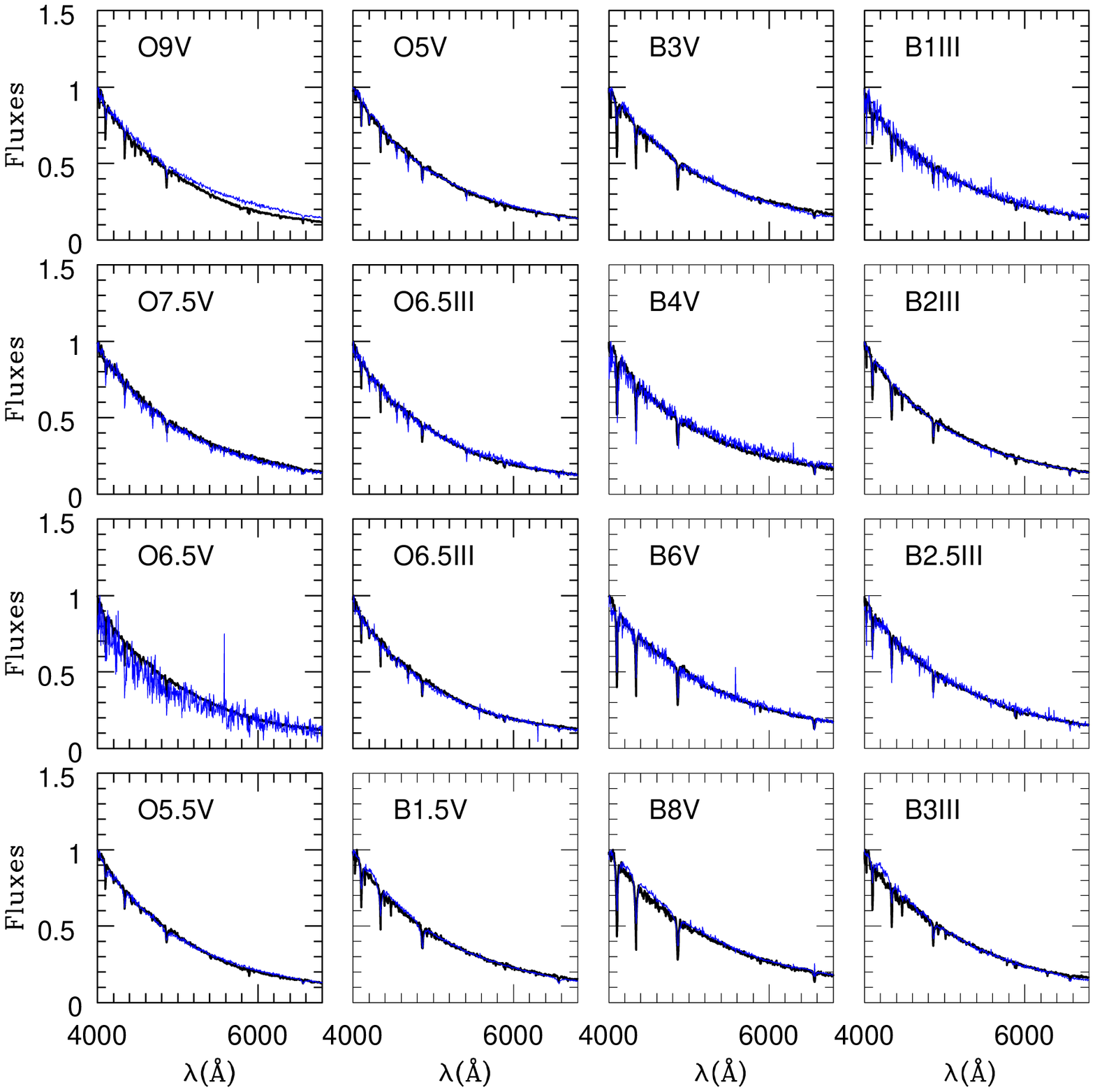}
\caption{SDSS-DR2 spectra with the corresponding JHC reference
library spectra and its spectro-luminosity type (Table 1).}
\end{figure}
%%%
\clearpage
\begin{figure}
\hskip -1 cm
\includegraphics{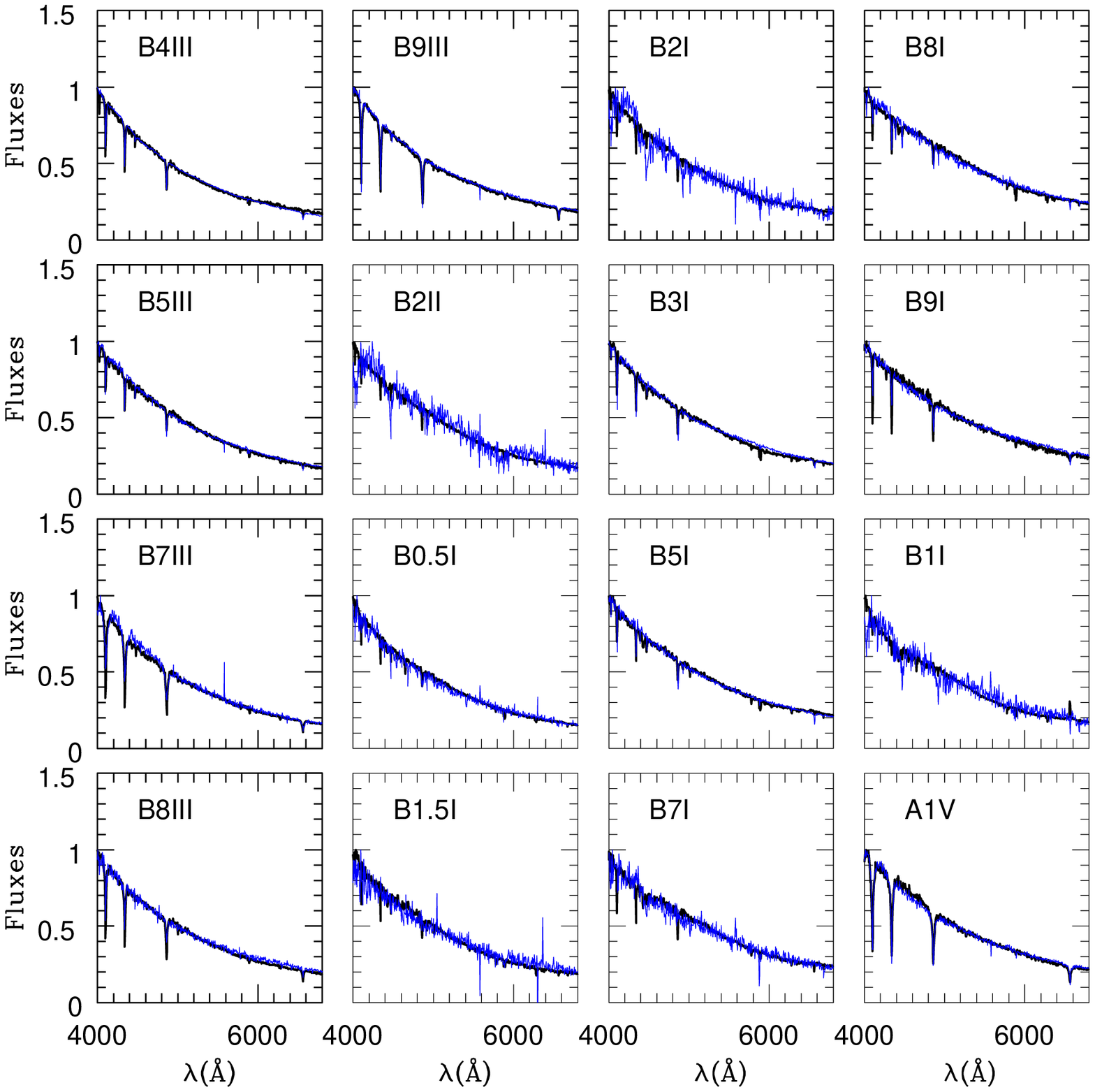}
\caption{SDSS-DR2 spectra with the corresponding JHC reference
library spectra and its spectro-luminosity type (Table 1).}
\end{figure}
%%%
\clearpage
\begin{figure}
\hskip -1 cm
\includegraphics{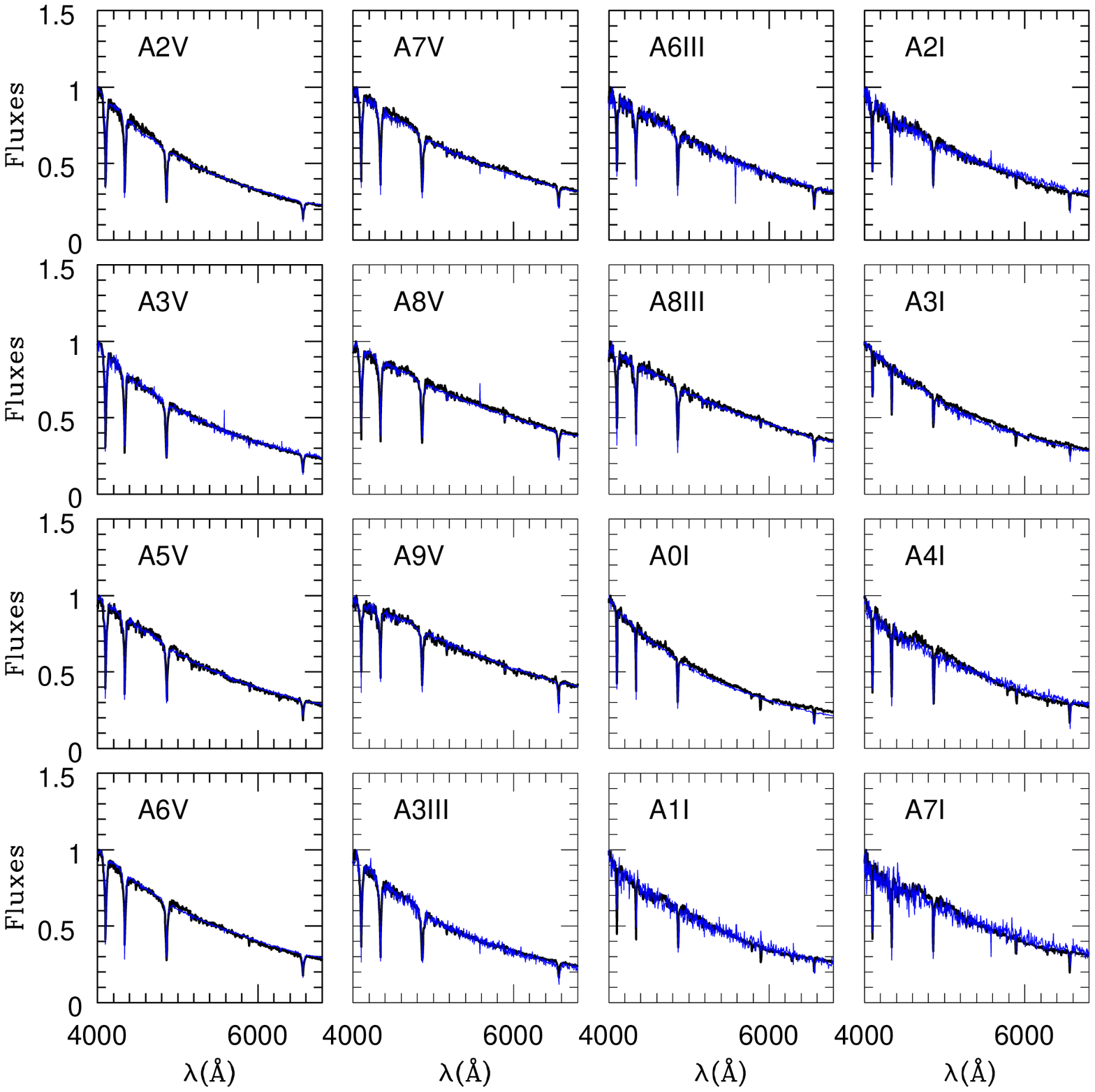}
\caption{SDSS-DR2 spectra with the corresponding JHC reference
library spectra and its spectro-luminosity type (Table 1).}
\end{figure}
%%%
\clearpage
\begin{figure}
\hskip -1 cm
\includegraphics{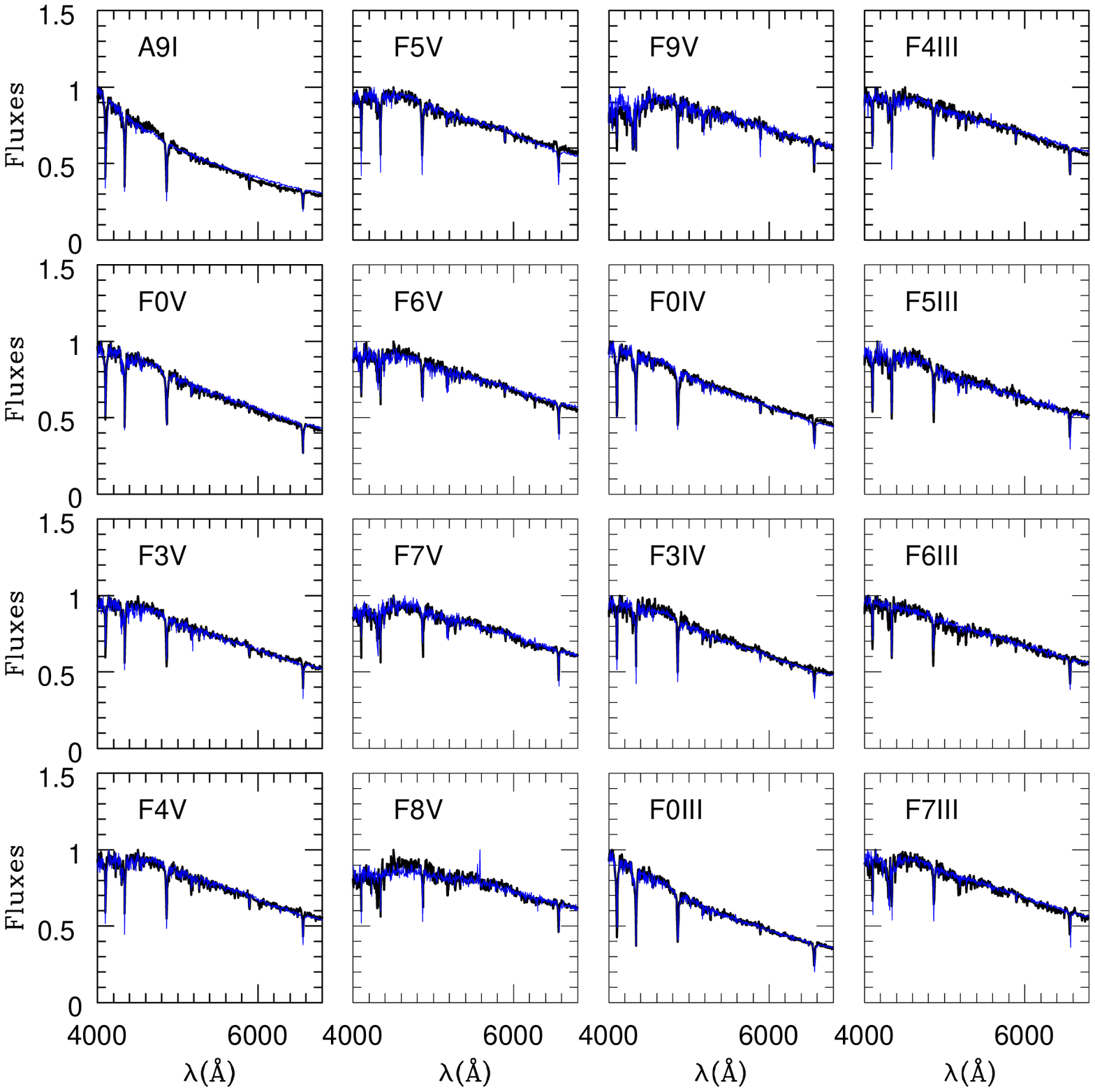}
\caption{SDSS-DR2 spectra with the corresponding JHC reference
library spectra and its spectro-luminosity type (Table 1).}
\end{figure}
%%%
\clearpage
\begin{figure}
\hskip -1 cm
\includegraphics{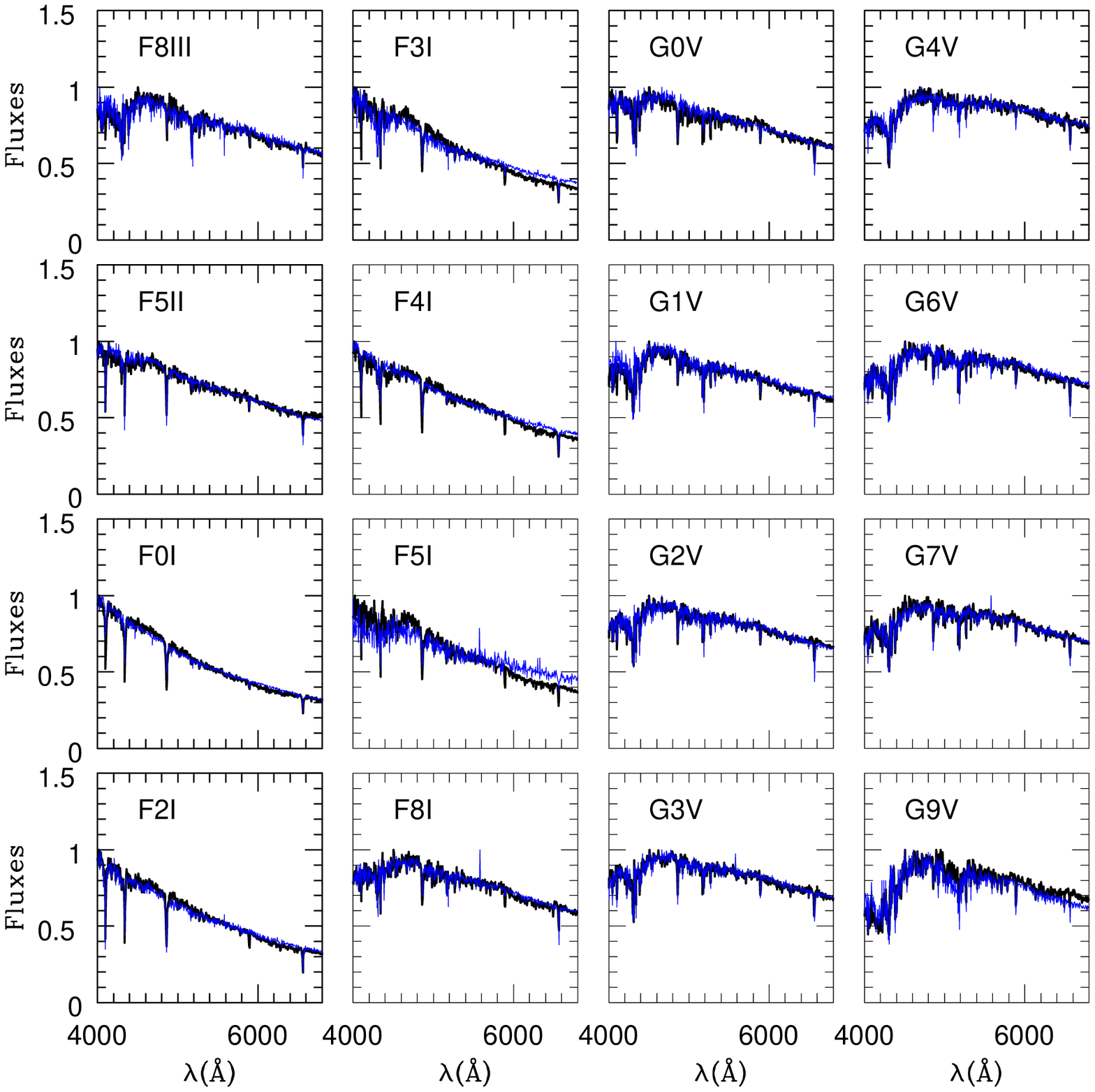}
\caption{SDSS-DR2 spectra with the corresponding JHC reference
library spectra and its spectro-luminosity type (Table 2).}
\end{figure}
%%%
\clearpage
\begin{figure}
\hskip -1 cm
\includegraphics{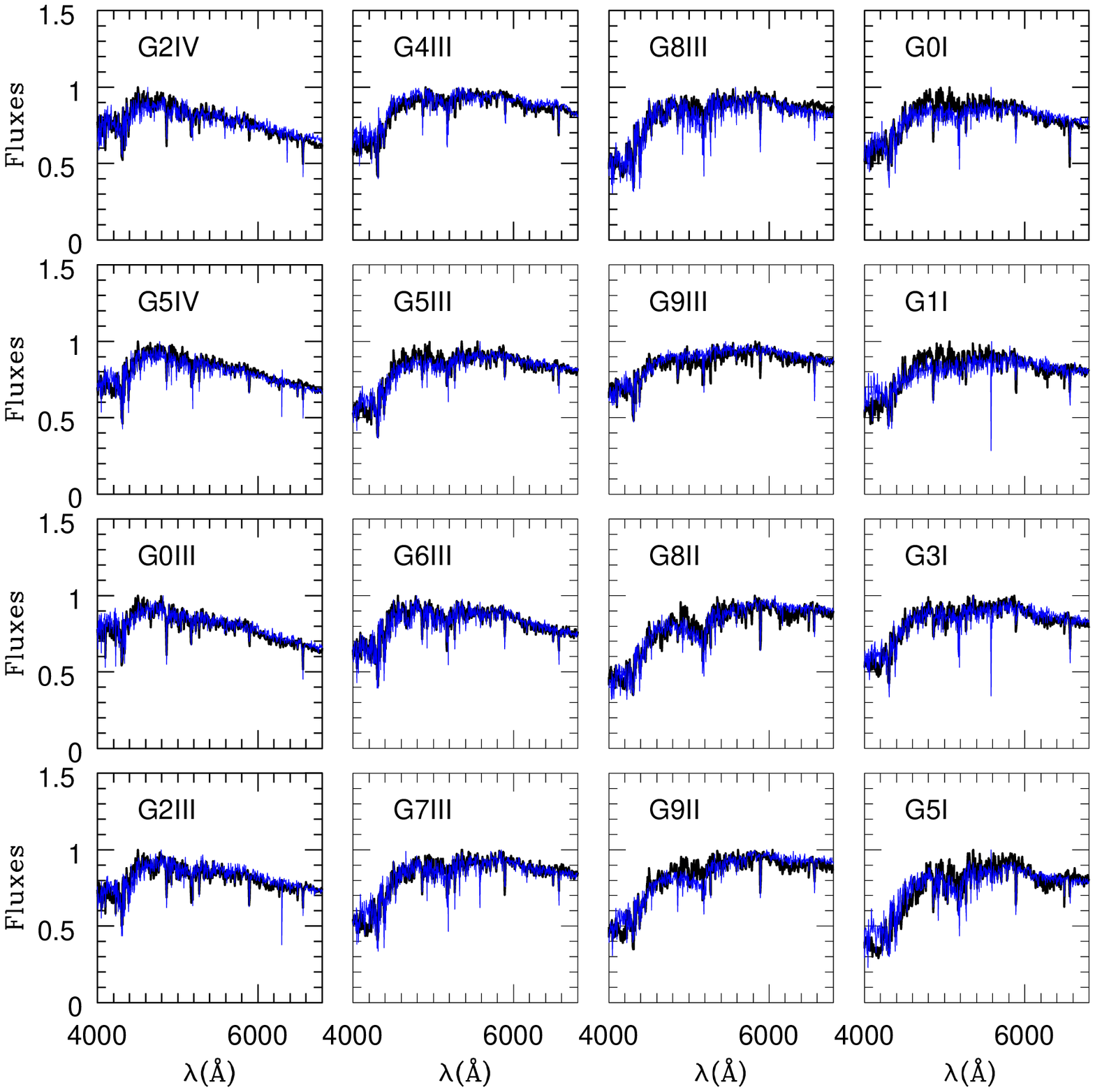}
\caption{SDSS-DR2 spectra with the corresponding JHC reference
library spectra and its spectro-luminosity type (Table 2).}
\end{figure}
%%%
\clearpage
\begin{figure}
\hskip -1 cm
\includegraphics{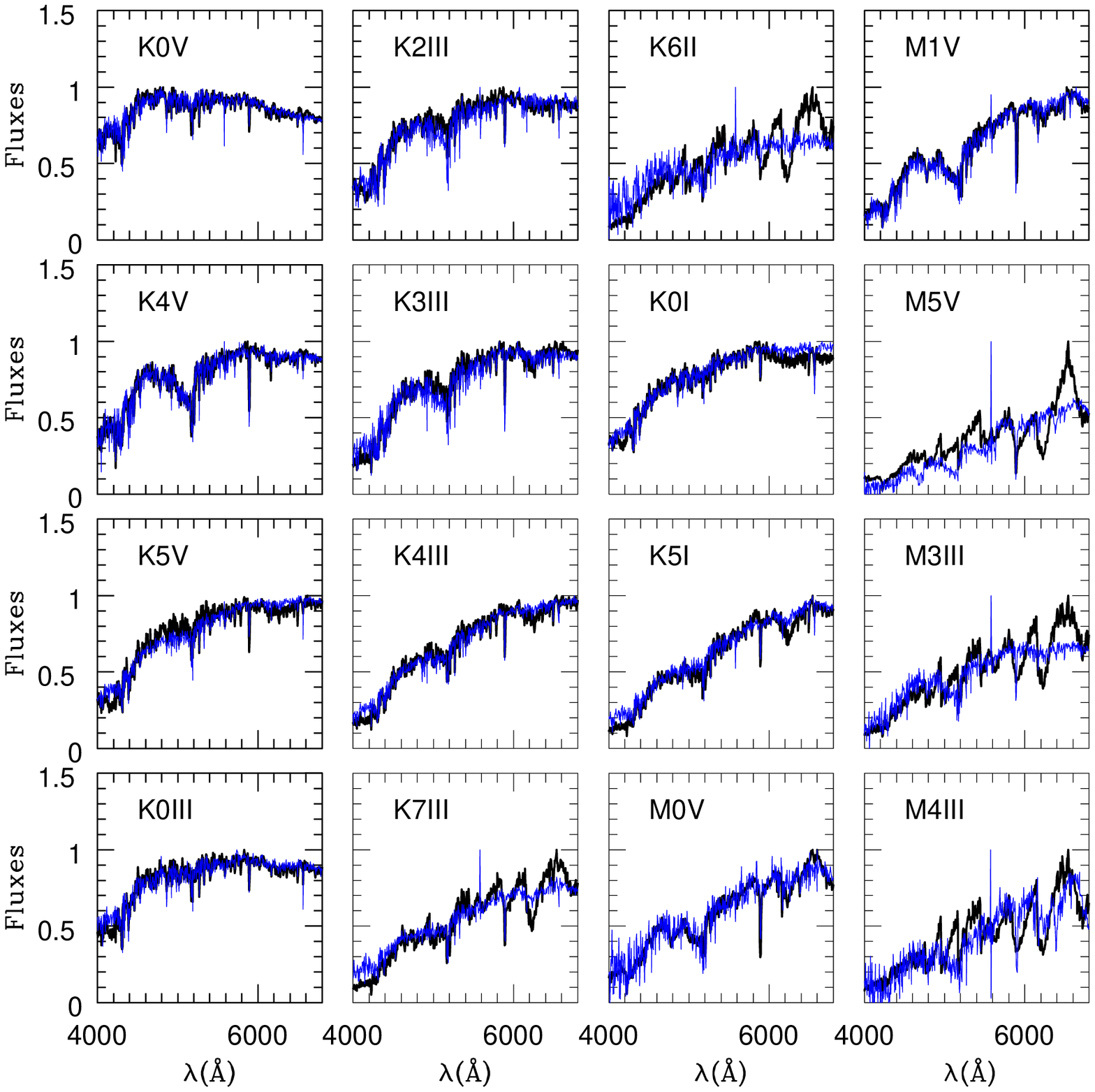}
\caption{SDSS-DR2 spectra with the corresponding JHC reference
library spectra and its spectro-luminosity type (Table 2).}
\end{figure}
%%%
\clearpage
\begin{figure}
\hskip -1 cm
\includegraphics{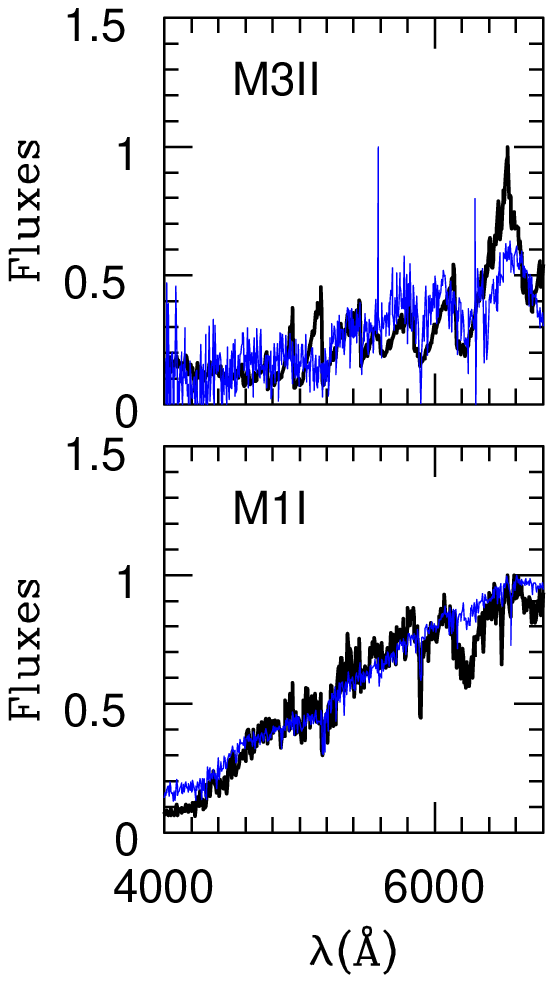}
\caption{SDSS-DR2 spectra with the corresponding JHC reference
library spectra and its spectro-luminosity type (Table 2).}
\end{figure}
%%%
%%%
\clearpage
\begin{figure}
\hskip -1 cm
\includegraphics{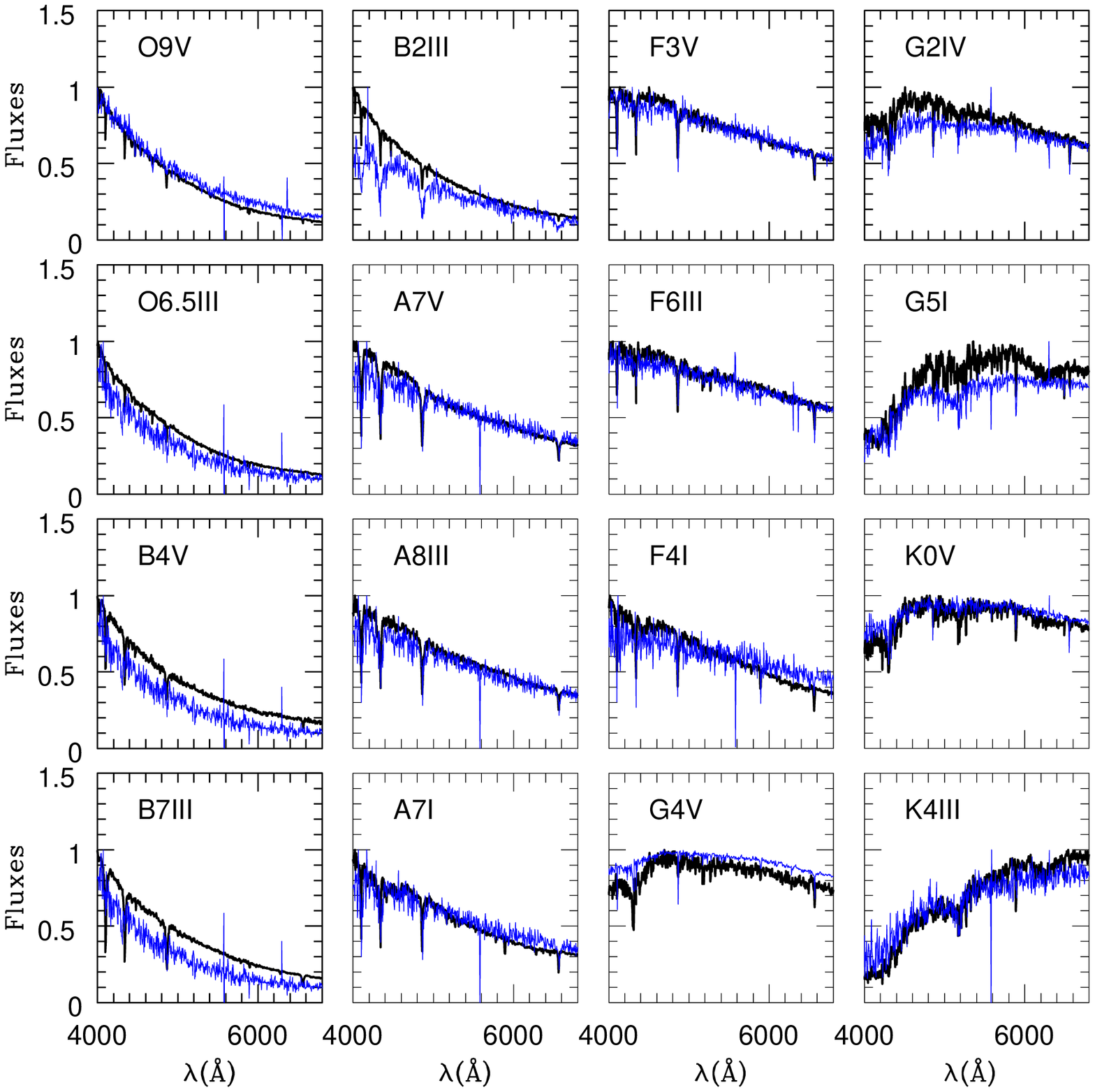}
\caption{SDSS-DR2 spectra with the corresponding JHC reference
library spectra and its spectro-luminosity type (Table 3).}
\end{figure}
%%%
\clearpage
\begin{figure}
\hskip -1 cm
\includegraphics{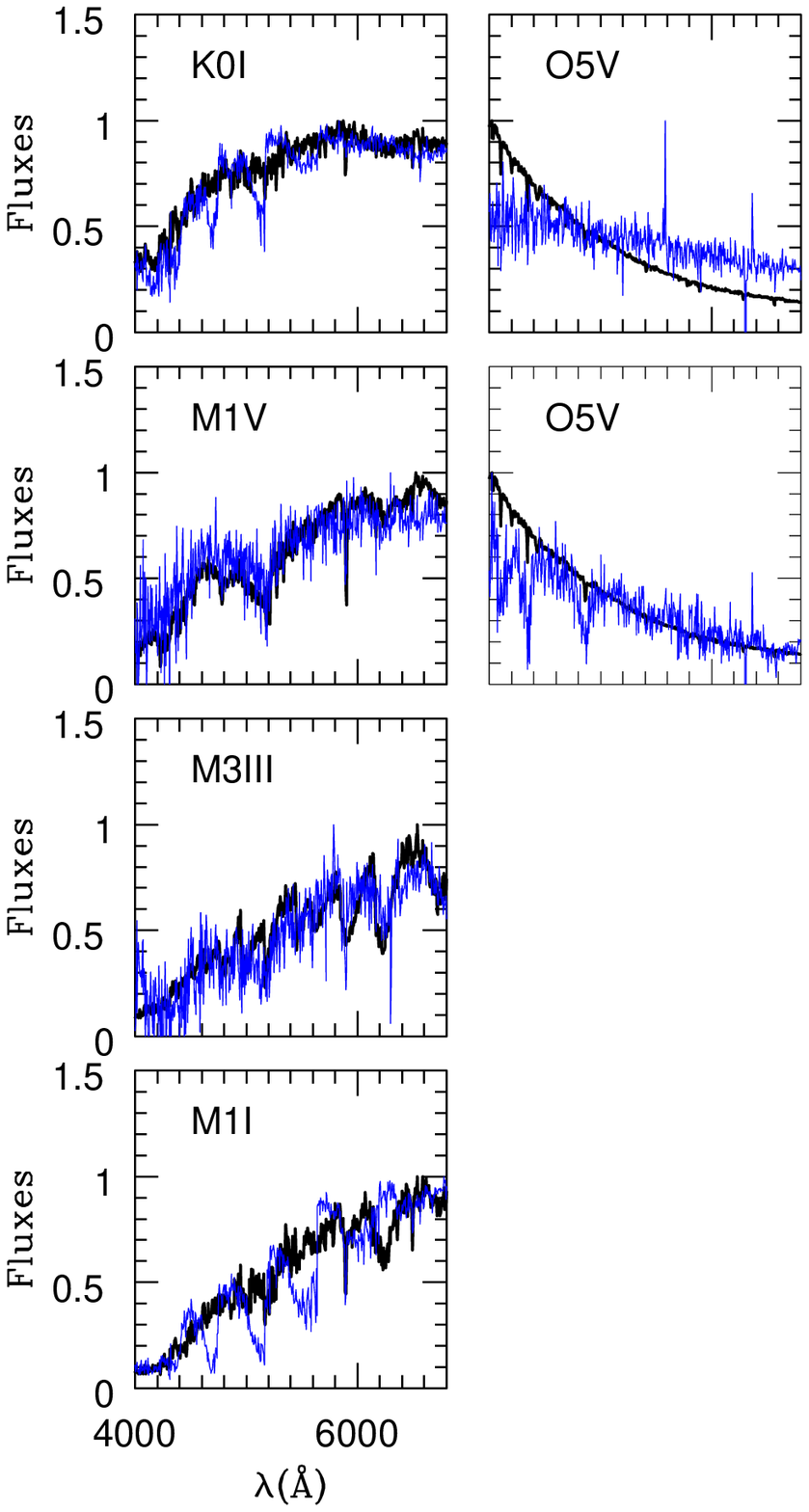}
\caption{SDSS-DR2 spectra with the corresponding JHC reference
library spectra and its spectro-luminosity type (Table 3).}
\end{figure}


\begin{thebibliography}{99}
\bibitem[Abazajian K. et al.(2003)]{b1} Abazajian K. et al., 2003, AJ, 126, 2081
\bibitem[Abazajian K. et al.(2004)]{b2} Abazajian K. et al., 2004, AJ, 128, 502
\bibitem[Abazajian K. et al.(2005)]{b3} Abazajian K. et al., 2005, AJ, 129, 1755
\bibitem[J. Adelman-McCarthy et al.(2006)]{b21} Adelman-McCarthy J. et al., 2006, ApJS 162, 38
\bibitem[J. Adelman-McCarthy et al.(2007a)]{b22} Adelman-McCarthy J. et al., 2007a, ApJS in press.
\bibitem[J. Adelman-McCarthy et al.(2007b)]{b23} Adelman-McCarthy J. et al., 2007b, Submitte to ApJS.
\bibitem[Allende Prieto, C.(2004)]{Allende}
Allende Prieto, C. 2004, AN, {\bf 325}, No. 6-8, 604
\bibitem[N. M. Ball et al.(2004)]{b12} N. M. Ball, J. Loveday, M. Fukugita, O. Nakamura, S. Okamura, J. Brinkmann and R. J. Brunner 2004, MNRAS, 348, 1038
\bibitem[Gulati et al.(1994)]{b7}Gulati R. K., Gupta R., Khobragade S., 1994, ApJ, 426, 340
\bibitem[Gulati et al.(1994)]{b8} Gulati R. K., Gupta R., Khobragade S., 1994, Vistas in Astronomy, 38, 293
\bibitem[Gulati et al.(1995)]{b9} Gulati R. K.,Gupta R., Khobragade S., 1995, Astronomica Data Analysis Software and Systems IV, ASP Conf. Ser. vol. 77, R. A. Shaw, H.E. Payne,and J.J.E. Hayes, eds. p.253
\bibitem[Gupta et al.(2004)]{b10} Gupta Ranjan, Singh Harinder P., Volk K., Kwok S., 2004, ApJS, 152, 201
\bibitem[Jacoby et al.(1984)]{b11} Jacoby G. H., Hunter D. A., Christian C. A. 1984, ApJS, 56, 257(JHC).
\bibitem[E. Parzen(1962)]{b6} E. Parzen 1962, Annals of Mathematical Statistics, 3, 1065
\bibitem[Qin, Dong-Mei et al.(2003)]{b13} Qin Dong-Mei, Guo Ping, Hu Zhan-Yi and Zhao Yong-Heng 2003, Chinese Journal of Astronomy \& Astrophysics, 3, 277
\bibitem[Singh et al.(1998)]{b14} Singh Harinder P., Gulati Ravi K., Gupta Ranjan. 1998, MNRAS, 295, 312
\bibitem[Singh et al.(2003)]{b15} Singh Harinder P., Gupta Ranjan. 2003, Large Telescopes and Virtual Observatory: Visions for the Future, 25th meeting of the IAU, Joint Discussion 8.
\bibitem[D. F. Specht(1990)]{b4} Specht D. F., 1990, Neural Network, 1(3), p.109
\bibitem[Specht \& Romsdahl(1994)]{b5} Specht  D. F., Romsdahl H., 1994, In Proceedings of the IEEE International Conference on Neural Networks, vol. 2, p.1203
\bibitem[Stoughton et al.(2002)]{b16} Stoughton, Chris. 2002, AJ, 123, 3487
\bibitem[Vanzella, E. et al.(2004)]{b17} Vanzella E. et al. 2004, A\&A, 423, 761
\bibitem[Von Hippel et al.(1994)]{b18} Von Hipple T., Storrie-Lombardi L. J., Storrie-Lombardi M. C., Irwin M. J. 1994, MNRAS, 269, 97
\bibitem[Weaver et al.(1995)]{b19} Weaver Wm. Bruce, Torres-Dodgen Ana V. 1995, ApJ,  446, 300
\bibitem[York et al.(2000)]{b20} York D.G., Adelman J., Anderson J. E. et al., 2000, AJ, 120, 1579

\end{thebibliography}
\end{document}